\documentclass[prl,twocolumn,amsmath,amssymb,superscriptaddress]{revtex4-1}

\usepackage{graphicx}
\usepackage{dcolumn}
\usepackage{hhline}
\usepackage{bm,color}

\usepackage{float}

\usepackage{amsmath}

\usepackage{amsfonts}

\usepackage{amssymb}

\usepackage{epstopdf}

\usepackage{multirow}

\begin{document}


\title{Exotic low-energy excitations emergent in the random Kitaev magnet Cu$_2$IrO$_3$}

\author{Y. S. Choi}
\affiliation{Department of Physics, Chung-Ang University, Seoul 06974, Republic of Korea}

\author{C. H. Lee}
\affiliation{Department of Physics, Chung-Ang University, Seoul 06974, Republic of Korea}

\author{S. Lee}
\affiliation{Department of Physics, Chung-Ang University, Seoul 06974, Republic of Korea}

\author{Sungwon Yoon}
\affiliation{Department of Physics, Chung-Ang University, Seoul 06974, Republic of Korea}

\author{W.-J. Lee}
\affiliation{Department of Physics, Chung-Ang University, Seoul 06974, Republic of Korea}

\author{J. Park}
\affiliation{Department of Physics, Chung-Ang University, Seoul 06974, Republic of Korea}

\author{Anzar Ali}
\affiliation{Department of Physical Sciences, Indian Institute of Science Education and Research Mohali,
Sector 81, S. A. S. Nagar, Manauli 140306, India}

\author{Yogesh Singh}
\affiliation{Department of Physical Sciences, Indian Institute of Science Education and Research Mohali,
Sector 81, S. A. S. Nagar, Manauli 140306, India}

\author{Jean-Christophe Orain}
\affiliation{Laboratory for Muon Spin Spectroscopy, Paul Scherrer Institute, 5232 Villigen PSI, Switzerland}

\author{Gareoung Kim}
\affiliation{Department of Applied Physics, Kyung Hee University, Yongin 17104, Republic of Korea}

\author{Jong-Soo Rhyee}
\affiliation{Department of Applied Physics, Kyung Hee University, Yongin 17104, Republic of Korea}

\author{Wei-Tin Chen}
\affiliation{Center for Condensed Matter Sciences, National Taiwan University, Taipei 10617, Taiwan}

\author{Fangcheng Chou}
\affiliation{Center for Condensed Matter Sciences, National Taiwan University, Taipei 10617, Taiwan}
\affiliation{National Synchrotron Radiation Research Center, Hsinchu 30076, Taiwan}
\affiliation{Taiwan Consortium of Emergent Crystalline Materials, Ministry of Science and Technology, Taipei 10622, Taiwan}

\author{Kwang-Yong Choi}
\email[]{kchoi@cau.ac.kr}
\affiliation{Department of Physics, Chung-Ang University, Seoul 06974, Republic of Korea}

\begin{abstract}
We report on magnetization $M(H)$, dc/ac magnetic susceptibility $\chi(T)$, specific heat $C_{\mathrm{m}}(T)$ and muon spin relaxation ($\mu$SR) measurements of the Kitaev honeycomb iridate Cu$_2$IrO$_2$ with quenched disorder. In spite of the chemical disorders, we find no indication of spin glass down to 260~mK from the $C_{\mathrm{m}}(T)$ and $\mu$SR data. Furthermore, a persistent spin dynamics observed by the zero-field muon spin relaxation evidences an absence of static magnetism. The remarkable observation is a scaling relation of $\chi[H,T]$ and $M[H,T]$ in $H/T$ with the scaling exponent $\alpha=0.26-0.28$, expected from bond randomness. However, $C_{\mathrm{m}}[H,T]/T$ disobeys the predicted universal scaling law, pointing towards the presence of low-lying excitations in addition to random singlets. Our results signify an intriguing role of quenched disorder in a Kitaev spin system in creating low-energy excitations possibly pertaining to Z$_2$ fluxes.

\end{abstract}

\maketitle
The exactly solvable Kitaev  honeycomb model provides a novel route to achieve elusive topological and quantum spin liquids~\cite{Kitaev,Baskaran}. Exchange frustration of bond-dependent Ising interactions fractionalizes the $j_\mathrm{eff}=\frac{1}{2}$ spin into itinerant Majorana fermion and static Z$_2$ gauge flux~\cite{Knolle13,Knolle14,Nasu}. Edge-sharing of octahedrally coordinated metal ions subject to strong spin-orbit coupling supports the realization of Kitaev-type interactions~\cite{Jackeli,Chaloupka,Jeffrey}.

In the quest for a Kitaev honeycomb magnet, the family of A$_2$IrO$_3$ (A = Na, Li) and  $\alpha$-RuCl$_3$ are considered prime candidate materials~\cite{Singh,Choi,Takayama,Modic,Glamazda,Plumb,Banerjee,Do}.  In these compounds, however, the theoretically predicted spin-liquid state is preempted by long-range magnetic order due to structural imperfections. As the real materials are vulnerable to a monoclinic stacking of honeycomb layers, non-Kitaev terms seem inevitable. A related issue is to engineer local crystal environments towards an optimal geometry to maximize the Kitaev interactions.

Very recently, the new Kitaev honeycomb iridates H$_3$LiIr$_2$O$_6$ and Cu$_2$IrO$_3$ have been derived from their ancestors A$_2$IrO$_3$ through soft structural modifications~\cite{Kitagawa,Abramchuk}. H$_3$LiIr$_2$O$_6$ is obtained by replacing the interlayer Li$^+$ ions with H$^+$ from $\alpha$-Li$_2$IrO$_3$, while the honeycomb layer remains intact. A scaling of the specific heat and NMR relaxation rate gives evidence for the presence of fermionic  excitations~\cite{Kitagawa}. In stabilizing a Kitaev-like spin liquid, hydrogen disorders turn out to a key ingredient by enhancing  Kitaev exchange interactions and promoting spin  disordering~\cite{Slagle,Yadav}. In case of Cu$_2$IrO$_3$, all of the A-site cations of Na$_2$IrO$_3$ are permuted by Cu$^+$ ions. Consequently, in-plane bond disorders become significant in determining magnetic behavior.

\begin{figure*}
	\centering
	\includegraphics[width=0.95\linewidth]{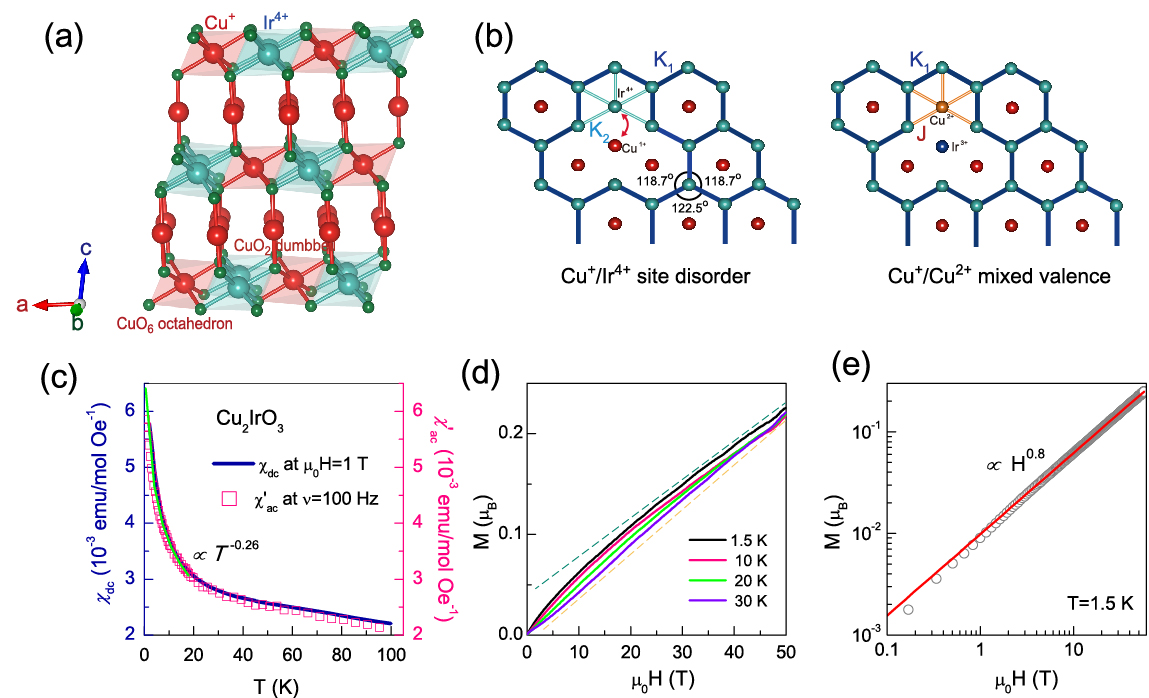}
	\caption{(a) Crystal structure of Cu$_2$IrO$_3$ consisting of edge-sharing
(Ir$_{2/3}$Cu$_{1/3}$)O$_6$ octahedra in the honeycomb layers and linear CuO$_2$ dumbbells between the layers. The greenish blue, red, and green spheres represent Ir, Cu and O ions, respectively.
 (b) Sketch of two types of chemical disorders occurring in the honeycomb plane. Cu$^{1+}$/Ir$^{4+}$ cation disorders 
 and Cu$^{1+}$(Ir$^{4+}$)/Cu$^{2+}$(Ir$^{3+}$) mixed valence  not only create nonmagnetic impurities but also supply a new triangular motif.
  (c) Comparison of the dc and ac magnetic susceptibility of Cu$_2$IrO$_3$ measured with an applied field of 1~T and at $\nu=100$~Hz and $H_{\mathrm{osc}}=10$~Oe, respectively. The green solid line is a fit to
  a power-law $\chi(T)\sim T^{-\alpha_s}$ with $\alpha_s=0.26$.
  (d)  High-field magnetization curves $M(H)$ measured at $T=1.5, 10, 20, \,\mbox{and}\, 30$~K. (e) Log-log plot of $M(H)$ at $T=1.5$~K.  The solid red line is a power-law fit
         $M(H)\sim H^{1-\alpha_m}$ with $\alpha_m\approx 0.2$.}
	\label{fig1}
\end{figure*}

Figure~\ref{fig1}(a) presents the crystal structure of Cu$_2$IrO$_3$ (isostructural to Na$_2$IrO$_3$ with {\it C2/c} space group), in which the honeycomb layers are stacked by CuO$_2$ dumbbells, distinct from CuO$_6$ octahedra within the honeycomb layers. The interlayer dumbbell structure arises from the eclipsed stacking of adjacent layers that align the oxygen and interlayer copper atoms in a line. We note that the CuO$_2$ dumbbells in Cu$_2$IrO$_3$ and the linear O-H-O links in H$_3$LiIr$_2$O$_6$ have structural similarities in bridging the stacking of honeycomb layers. This structural alteration  leads to an elongation of the $c$-axis and closeness of the Ir-Ir-Ir bond angles to the ideal 120$^{\circ}$, compared to its predecessor  Na$_2$IrO$_3$ [see Fig.~\ref{fig1}(b)]. However, little is known about its underlying magnetism.

In this paper, we provide thermodynamic and $\mu$SR spectroscopic signatures of a proximate spin-liquid state. We find a scaling relation of the $(T,H)$-dependent magnetization and $ac$ susceptibility in $T/H$, but not of the $(T,H)$-dependent specific heat. This observation suggests the presence of emergent low-lying excitations in addition to the random singlets expected from bond disorders.

Polycrystalline samples of  Cu$_2$IrO$_3$  were prepared by a topotactic reaction
as described in Ref.~\cite{Abramchuk}.
Dc and ac magnetic susceptibilities were measured using a SQUID and  a vibrating sample magnetometer (Quantum Design MPMS and VSM). The magnetization measurements up to 14~T  were carried out using a Physical Property Measurement System (Quantum Design PPMS Dynacool). Specific heat was measured with the thermal-relaxation method using an option of the PPMS apparatus with  a $^3$He insert. High-field magnetization curves were recorded with a nondestructive pulsed magnet  at the Dresden High Magnetic Field Laboratory.
Zero-field (ZF)- and longitudinal-field (LF)-$\mu$SR  experiments were performed with the DOLLY
spectrometer at PSI (Villigen, Switzerland). For measurements down to 0.26~K, the samples were loaded into the Variox cryostat equipped with Heliox insert.
All of the obtained $\mu$SR data were analyzed using the software package Musrfit~\cite{Suter}.

Figure~\ref{fig1}(c) shows the $T-$dependence of the static magnetic susceptibility $\chi_{\mathrm{dc}}(T)$ of Cu$_2$IrO$_3$ measured at $\mu_0H=1$~T, together with the real component of {\it ac} susceptibility
$\chi'_{\mathrm{ac}}(T)$. As $T\rightarrow 0$~K, both $\chi_{\mathrm{dc}}(T)$
and $\chi'_{\mathrm{ac}}(T)$ exhibit a steep increase without obvious saturation or kink, in spite
of the large Curie-Weiss temperature $\Theta_{\mathrm{CW}}=-110$~K~\cite{Abramchuk}. It is remarkable that  both $\chi_{\mathrm{dc}}(T)$ and $\chi'_{\mathrm{ac}}(T)$ are described by a power-law increase $\chi(T)\sim T^{-\alpha_s}$ with $\alpha_s=0.26$ for temperatures below 20~K. Such a sub-Curie law behavior is indicative of the presence of abundant low-energy excitations.
As shown in Fig.~\ref{fig1}(d), the high-field magnetization curves $M(H)$ at selected low temperatures display clear deviations from a linear-$H$ dependence. Significantly, we find that the $T=1.5$~K $M(H)$ follows a power-law dependence $M(H)\sim H^{1-\alpha_m}$ with $\alpha_m=0.2$ over an entirely measured field range [see a log-log plot of $M(H)$ in Fig.~\ref{fig1}(e)]. The commonly observed
power-law behavior of $\chi(T)$ and $M(H)$ with $\alpha_s\approx \alpha_m$ constitutes a hallmark of
random magnetic interactions~\cite{Ma,Hirsch,Kimchi}.

\begin{figure*}
	\centering
	\includegraphics[width=1.0\linewidth]{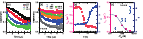}
	\caption{(a) ZF-$\mu$SR spectra of Cu$_2$IrO$_3$ at selected temperatures. (b) Field dependence of LF-$\mu$SR spectra measured at $T=0.26$~K in an applied  field of $H=0 - 4500$~G. The solid lines are fits to the stretched exponential function. (c) The muon relaxation rate $\lambda_{\mathrm{ZF}}(T)$ and the stretch exponent $\beta(T)$ extracted from fitting to the ZF data. (d) Longitudinal-field dependence of the muon relaxation rate $\lambda_{\mathrm{LF}}(H_{\mathrm{LF}})$ and the stretch exponent $\beta(H_{\mathrm{LF}})$  extracted from the $T=0.26$~K LF data. The solid curve denotes the fit of $\lambda_{\mathrm{LF}}(H_{\mathrm{LF}})$ to the Redfield formula.}
	\label{fig2}
\end{figure*}

Given the sizable Cu$^{1+}$/Ir$^{4+}$ intersite disorder and Cu$^{1+}$/Cu$^{2+}$ mixed valence~\cite{Abramchuk}, magnetic ions (either Ir$^{4+}$ or Cu$^{2+}$)  occupy randomly at the center of the honeycomb lattice, while nonmagnetic ions (either Cu$^{1+}$ or Ir$^{3+}$) go
on the vertex of the honeycomb lattice. Based on the power-law dependence of $M(H)$ persisting down to 0.3~T in Fig.~\ref{fig1}(e), we conclude that the chemical disorders are mostly confined within the honeycomb layers. As sketched in Fig.~\ref{fig1}(b), an impact of the quenched disorders on magnetism is twofold. One is to introduce
spin vacancies, and the other is to generate a spin triangular motif which has a different coupling constant from the original honeycomb lattice, thereby creating a random distribution  in exchange constants.

An essence of random magnetism consists in low-energy random singlets generated
in the matrix of a quantum disordered state, which give rise to a diverging
low-energy excitation~\cite{Ma,Hirsch}. Before proceeding, we first clarify whether the quenched disorder induces frozen moments.
For this purpose, we employ a $\mu$SR technique, allowing differentiating a quantum
disordered from a weakly ordered state.

Figures~\ref{fig2}(a) and~\ref{fig2}(b) show the ZF-$\mu$SR spectra at selected temperatures and the LF-$\mu$SR spectra measured
at $T =0.26$~K in an applied field of $H =0 - 4500$~G. The $\mu$SR data
at all temperatures and fields are successfully described by the stretched-exponential function
$P_{\mathrm{ZF}}(t)=P_0\exp[-(\lambda_{\mathrm{ZF}}t)^{\beta}]+P_{\mathrm{bg}}$, where $\lambda_{\mathrm{ZF}}$ is the muon relaxation rate and $\beta$ is the stretching
exponent. The $T$-independent $P_{\mathrm{bg}}$ term represents a background
contribution, which originates from muons implanted in the silver sample holder or the cryostat.

With decreasing temperature, the ZF-$\mu$SR spectra gradually change their asymmetry lineshape
from a Gaussian- to a Lorentzian-like relaxation form. At the measured lowest temperature
of 0.26~K, we observe neither spontaneously oscillating component
nor missing initial $P_{\mathrm{ZF}}(t)$. Rather, the low-$T$ ZF-$\mu$SR data display a dynamic relaxation with no indication to a recovery of a one-third tail at long times. The temperature evolution of a relaxation lineshape excludes any local static fields at the muon stopping site.
Also, the LF dependence of the $\mu$SR spectra supports that the Ir moments
remain in a dynamically fluctuating state at least down to $T=0.26$~K.
As plotted in Fig.~\ref{fig2}(b), even applying the longitudinal field of $H_{\mathrm{LF}}=4500$~G, there is a substantially decoupled relaxation.
If the low-$T$ muon spin relaxation at zero field arises from static magnetism, the local static field is estimated to be $B_{\mathrm{loc}}\approx 4.4$~G.
In this case, one would observe a full polarization of the $\mu$SR spectra
upon the application of about 50~G longitudinal fields.

\begin{figure*}
	\centering
	\includegraphics[width=0.95\linewidth]{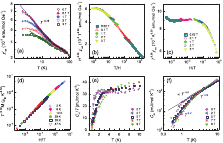}
	\caption{ (a) The real component of the ac susceptibility $\chi'_{\mathrm{ac}}(T)$
plotted against temperature for Cu$_2$IrO$_3$ under various external fields.
The solid line is a power-law dependence of the 0~T data, $\chi'_{\mathrm{ac}}(T)\sim T^{-0.26}$.
(b) Log-log scaled plot of $H^{0.26}\chi'_{\mathrm{ac}}$ vs $T/H$ displays a scaling relation.
(c) $T^{0.26}\chi'_{\mathrm{ac}}$ shows a scaling with $H/T$.  (c) $T-H$ scaling of $M(H)$ plotted on
a log-log scale. (e) Magnetic specific heat divided by temperature
$C_m/T$ versus $T$ for various magnetic fields. (f) Log-log plot of the same $(T,H)$-dependent $C_m$. The solid lines are the fits of the $C_m$ data to a power law.}
\label{fig3}
\end{figure*}

We next turn to the $T$ and $H$ evolution of the obtained fit parameters.
As seen from Fig.~\ref{fig2}(c), with decreasing temperature $\lambda_{\mathrm{ZF}}(T)$
starts to increase at 6~K and then saturates at about 1~K. Such a low-$T$ plateau  in $\lambda_{\mathrm{ZF}}(T)$, representing a persistent spin dynamics,
 has been observed in a range of geometrically frustrated magnets having spin freezing or weak magnetic order, let alone quantum spin liquids~\cite{Gardner,Mendels,Li,Yaouanc1}. In parallel to $\lambda_{\mathrm{ZF}}(T)$, the stretching exponent decreases rapidly
from $\beta>1$ to $\beta<1$ on cooling through 10~K, and then levels off to $\beta\approx 0.5$ below 2~K. The obtained value of $\beta$ at the base temperature  is larger than
$\beta=1/3$ (value of canonical spin glass).

Shown in Fig.~\ref{fig2}(d) is the LF-dependence of $\lambda_{\mathrm{LF}}(H_{\mathrm{LF}})$ and $\beta(H_{\mathrm{LF}})$ measured at $T=0.26$~K. In the Redfield  model, $\lambda_{\mathrm{LF}}(T)$
is associated with the fluctuation frequency $\nu$ and the fluctuating time-averaged local field
$\langle H_{\mathrm{loc}}^2\rangle$ by $\lambda_{\mathrm{LF}}(H_{\mathrm{LF}})=2\gamma_\mu^2 \langle H_{\mathrm{loc}}^2\rangle\nu/(\nu^2+\gamma_\mu^2 H_{\mathrm{LF}}^2)$~\cite{Redfield, Yaouanc2}.
Here, $\gamma_\mu$ is the muon gyromagnetic ratio. From fits of the $T=260$~mK data, we obtain
$\nu\sim 72$~MHz and $H_{\mathrm{loc}}\sim 25$~G. The determined $\nu$ ($H_{\mathrm{loc}}$) is roughly two times slower (larger) than $\nu\sim 150$~MHz and $H_{\mathrm{loc}}\sim 18$~G of the well-investigated spin-liquid material  ZnCu$_3$(OH)$_6$Cl$_2$~\cite{Mendels}. On the LF application, $\beta(H_{\mathrm{LF}})$ steadily increases, implying the gradual quenching of widely distributed fluctuating local fields. Taken the $\mu$SR data together, we conclude that
Cu$_2$IrO$_3$ features a dynamically fluctuating state. Based on the relatively small value of $\beta\approx 0.5$, however, we cannot totally exclude
some tendency to weak spin freezing, if any.

In the followings, we will discuss an intricate relation between quenched disorder and critical spin correlations.

In Fig.~\ref{fig3}(a), the $H$ and $T$ dependences of $\chi'_{\text{ac}}$  are plotted on the log-log scale in the applied fields of $\mu_{0}H=0- 3$~T.  With increasing field, $\chi'_{\text{ac}}(T)$ is systematically reduced, yet still maintains a power-law dependence, while changing the exponent
from $\alpha_s=0.26$ at 0~T to 0.13 to 3~T. In an attempt to corroborate  dynamic scaling behavior,
we plot $H^{\alpha_s}\chi'_{\text{ac}}$ vs $T/H$ and  $T^{\alpha_s}\chi'_{\text{ac}}$ vs $H/T$ in Figs.~\ref{fig3}(b) and \ref{fig3}(c), respectively. Strikingly, the ($H$,\,$T$)-dependent $\chi'_{\text{ac}}$ data overlap over three orders of magnitude with the same value of $\alpha_s=0.26$. We provide further supportive evidence for universal scaling behavior from testing the similar scaling behavior in the $M(H)$ data measured up to $\mu_{0}H=14$~T at various temperatures (2, 5, 10, 20, 30, and 47~K). We plot $MT^{\alpha_m-1}$  against $H/T$ in Fig.~\ref{fig3}(d), and obtain the exponent value of $\alpha_m=0.28$ that guarantees a sufficient data collapse onto a single scaling curve. Within the error bars,  the scaling exponent of $0.26-0.28$ agrees exceptionally well between the two distinct thermodynamic quantities. This is reminiscent of universal scaling observed
in a certain class of frustrated quantum magnets H$_3$LiIr$_2$O$_6$, LiZn$_2$Mo$_3$O$_8$, and ZnCu$_3$(OH)$_6$Cl$_2$, which commonly share proximate spin liquids and quenched disorder~\cite{Kimchi}.

We further explore the specific heat scaling. Figure~\ref{fig3}(e) shows a $C_\mathrm{m}/T$ versus $T$ plot at low temperatures of $T=0.5-10$~K under the external magnetic fields of $\mu_0H=0-9$~T.
The zero-field $C_\mathrm{m}/T$ data shows little variation with temperature down to 2~K and then
a rapid drop below 1~K. With increasing magnetic field, the $C_\mathrm{m}/T$ contribution below 4~K
is steadily suppressed. Unlike the afore-mentioned quantum disordered magnets with quenched disorder,
the zero-field $C_\mathrm{m}/T$ data lacks any upturn at extremely low temperatures~\cite{Kimchi}.
In addition, we could find no scaling and data collapse of $C_\mathrm{m}[T,H]/T$ in $T/H$ (not shown here).

To figure out the failure of the predicted power-law scaling within a random magnetism scenario, we replot
$C_m$ vs $T$ on a log-log scale in Fig.~\ref{fig3}(f). The magnetic specific heat
is described by $C_\mathrm{m}(T)\sim T^{\gamma}$. Obviously, the exponent of $\gamma\approx 1.07$
at $\mu_0H=0$~T defies the $T^{1-\alpha}$ scaling ($\alpha=0.26-0.28$) obtained from $M(H)$ and $\chi'_{\mathrm{ac}}$.
Rather, a nearly $T$-linear contribution is reminiscent of gapless spinon excitations reported in the spin-liquid candidate materials~\cite{WJL}.
Furthermore, the low-$T$ and high-$H$ $C_\mathrm{m}$ data display a $T^{2.41}$ dependence,
stronger than the $T^2$ dependence often reported in the frustrated magnets with the
weak bond disorder, obeying the power-law scaling law~\cite{Kimchi}. Admittedly, there
is some uncertainty in evaluating the lattice contribution. As the $\chi'[T,H]$ and $M[T,H]$ scalings
are limited above 2~K, however, the specific heat anomaly may be resolved by invoking
the chemical disorder that alters extremely low-energy
excitations pertaining to $Z_2$ fluxes.

To conclude, we combine $\mu$SR with thermodynamic measurements to unravel
a ground state and bond-randomness-induced scalings in the random Kitaev magnet Cu$_2$IrO$_2$.

The ZF- and LF-$\mu$SR data evidence the formation of a spin-liquid-like ground state, in which
the spins remain almost dynamic down to 260~mK. This sets up a setting
for the random singlet model on the background of a quantum disordered state.
As anticipated, both magnetization and ac magnetic susceptibility obey a temperature-field scaling.
Thus, Cu$_2$IrO$_2$ seems to
belong to a class of frustrated quantum magnets that are proximate to a quantum disordered
state and are subject to quenched disorder. However, the purported random singlet scenario is not compatible with a lacking scaling in the magnetic specific heat. This highlights
an intriguing character of chemical disorders, which generates not only random exchange interactions but also spin vacancies
 [see Fig.~\ref{fig1}(b)]. Noteworthy is that the magnetic specific
heat shows a nearly $T$-linear relationship at low temperatures of $T=1-10$~K, typical for
gapless excitations, and then a strong $T$-dependence below 1~K, indicative of the development
of gapful excitations at extremely low temperatures.

Recent theoretical calculations show
that vison and Majorana zero mode can be created in the vicinity of a site vacancy~\cite{Udagawa}.
These zero-energy resonances are largely hidden to static spin susceptibility and, thus,
may provide a possible account of the specific heat anomaly.  Ultralow-temperature experiments
and future theoretical studies are requested to clarify whether this scenario is applicable
to Cu$_2$IrO$_2$.

{\it Note added.-} During writing the manuscript, we noted that a related study by E. M. Kenney {\it et al.}
was posted in preprint~\cite{Kenney}, which observed a coexistence of static and dynamic magnetism.
 Unlike their $\mu$SR data, the absence of frozen magnetic moments is clear in our sample, thereby
ruling out a nucleation of the quenched disorder.  As to the dynamic magnetism,
our data are consistent with their results.

\section*{ACKNOWLEDGMENTS}
 This work was supported by Korea Research Foundation (KRF) Grants (No. 2018-0189 and No. 2018-0099) funded by the Korea government (MEST). We acknowledge the support of the HLD at HZDR, member of the European Magnetic Field Laboratory (EMFL).

\bibliography{achemso}

\end{document}